\documentclass[12pt]{article}
\usepackage[english]{babel}
\usepackage{graphicx,epsfig}
\makeindex \textwidth 170mm \textheight 220mm \topmargin -5mm
\newcommand{\be}{\begin{equation}}
\newcommand{\ee}{\end{equation}}
\newcommand{\bea}{\begin{eqnarray}}
\newcommand{\eea}{\end{eqnarray}}

\def\rfr#1{eq.(\ref{#1})}
\def\rfrs#1#2{eqs.(\ref{#1})-(\ref{#2})}

\def\Rfr#1{Eq.(\ref{#1})}

\def\bb{\bibitem}
\def\eqi{\begin{equation}}
\def\eqf{\end{equation}}
\def\eqia{\begin{eqnarray}}
\def\eqfa{\end{eqnarray}}

\def\rp#1#2{{#1\over#2}}

\def\lb#1{\label{#1}}

\def\d{\delta}

\def\p{\pi}

\def\D{\Delta}

\begin{document}
\begin{center}
{\LARGE On the possibility of testing the Weak Equivalence
Principle with artificial Earth satellites} \vspace{1.0cm}
\quad\\
{Lorenzo Iorio$^{\dag}$\\
\vspace{0.1cm}
\quad\\
{\dag}Dipartimento di Fisica dell' Universit{\`{a}} di Bari, via
Amendola 173, 70126, Bari, Italy\\ \vspace{0.1cm} }
\quad\\
 \vspace{1.0cm}

{\bf Abstract\\}
\end{center}

{\noindent \footnotesize In this paper we examine the possibility
of testing the equivalence principle, in its weak form, by
analyzing the orbital motion of a pair of artificial satellites of
different composition moving along orbits of identical shape and
size in the gravitational field of Earth. It turns out that the
obtainable level of accuracy is, realistically, of the order of
$10^{-10}$ or slightly better. It is limited mainly by the fact
that, due to the unavoidable orbital injection errors, it would
not be possible to insert the satellites in orbits with exactly
the same radius and that such difference could be known only with
a finite precision. The present--day level of accuracy, obtained
with torsion balance Earth--based measurements and the analysis of
Earth--Moon motion in the gravitational field of Sun with the
Lunar Laser Ranging technique, is of the order of $10^{-13}$. The
proposed space--based missions STEP, $\mu$SCOPE, GG and SEE aim to
reach a $10^{-15}$--$10^{-18}$ precision level.}
%\end{titlepage}

\newpage \pagestyle{myheadings} \setcounter{page}{1}
\vspace{0.2cm} \baselineskip 14pt

\setcounter{footnote}{0}
\setlength{\baselineskip}{1.5\baselineskip}
\renewcommand{\theequation}{\mbox{$\arabic{equation}$}}
\noindent \section{Introduction} The Weak Equivalence Principle
(WEP) and the Einstein Equivalence Prnciple (EEP) (Will 1993;
Ciufolini and Wheeler 1995; Haugan and L${\rm \ddot a}$mmerzahl
2001; Will 2001) are the cornerstones of Einstein General Theory
of Relativity (GTR) and of all the other competing metric theories
of gravity. The WEP states that non--rotating, uncharged bodies of
different structure and compositions and with negligible amount of
gravitational binding energy per unit mass fall with the same
acceleration in a given gravitational field, provided that no
other forces act on them. The EEP (called also Medium Strong
Equivalence Principle) states that the outcome of any local,
non--gravitational test experiment is independent of where and
when in the gravitational field the experiment is performed. If
the gravitational binding energy per unit mass of the freely
falling bodies in a given external gravitational field is not
negligible, as is the case for astronomical bodies, we have the
Strong Equivalence Priciple (SEP. It is called also Very Strong
Equivalence Principle). When the self--gravity of the falling
bodies is accounted for it might happen, in principle, that their
accelerations are different. This violation of the universality of
free fall, called Nordvedt effect (Nordvedt 1968a), is predicted
by all metric theories of gravity apart from GTR which, instead,
satisfies the SEP. Lunar Laser Ranging (LLR) is able to test this
effect for the motion of Earth and Moon in the gravitational field
of Sun (Nordvedt 1968b). No different accelerations have been
found at a 10$^{-13}$ level (Anderson and Williams 2001). Of
course, this is also a test of the WEP.

To go to space, where much larger distances, velocities,
gravitational potential differences with respect to Earth, and,
most important, free fall for an, in principle, infinitely long
time are available, is important to perform very accurate tests of
post--Newtonian gravity. In this paper we wish to investigate the
level of accuracy which could be reached in testing the WEP, by
analysing the orbital motion of a pair of Earth artificial
satellites of different compositions for which the effects of
self--gravity would be, of course, negligible: it is analogous to
the analysis of the motion of Earth and Moon in the gravitational
field of Sun in order to test the SEP. For the already performed
experimental tests of the WEP on Earth see (Will 2001) and the
references therein. The present--day level of accuracy is $4\times
10^{-13}$ (B${\rm \ddot a}$ssler $et \ al$ 1999). It has been
obtained in the experiments of the so--called E${\rm \ddot
o}$t-Wash group of the Washington University by means of a
sophisticated torsion balance in an Earth based laboratory
set--up. Other proposed space--based experiments are the very
complex and expensive STEP (Lockerbie $et\ al$ 2001), $\mu$SCOPE
(Touboul 2001), GG (Nobili $et \ al$ 2000) and SEE (Sanders $et\
al$ 2000) missions\footnote{See also on the WEB
http://einstein.stanford.edu/STEP/,
http://www.onera.fr/microscope/
http://tycho.dm.unipi.it/$\sim$nobili/ggproject.html, and
http://www.phys.utk.edu/see/. Notice that while the first three
missions are in advanced stages of planning and hardware testing,
and are expected to be launched in the next few years, SEE is
still undergoing rigorous conceptual evaluation and is not yet a
scheduled mission.} whose goal is to reach the
$10^{-15}$--$10^{-18}$ level of accuracy.

The reason for searching for more and more accurate tests of the
equivalence principle resides in the fact that all approaches to
quantizing gravity and to unifying it with the other fundamental
interactions currently under study are capable of predicting
violations of the equivalence principle at some level. For
example, departures from universal free fall accelerations of the
order of $10^{-15}$ have been calculated in (Damour and Polyakov
1994a; 1994b) in the string theory context. Violations of the WEP
are predicted also by nonsymmetric theories of gravity (Will
1989). In (Moffat and Gillies 2002) a violation of the WEP of the
order of 10$^{-14}$ is predicted in the context of a non--local
quantum gravity theory.
%---------------------------------------------------------------------
\section{The orbital period}
In this section we examine the possibility of testing the WEP by
measuring the orbital periods of a pair of Earth artificial
satellites of different compositions. It can be thought of as a
comparison between two pendulums with enormously long threads
swinging for an extremely long time\footnote{In the case of the
SEP--LLR experiment the amplitude of the parallactic inequality
long--period harmonic perturbation, proportional to $\cos D$,
where $D$ is the synodic phase from New Moon, is sensitive to the
possible different falling rates of Earth and Moon toward the
Sun.}.

The orbital period of a satellite freely falling in Earth's
gravitational field can be written as \eqi T=\rp{2\pi}{n(1+\D
n)}\sim \rp{2\pi}{n}(1-\D n),\lb{uno}\eqf in which the Keplerian
unperturbed period is \eqi
T^{(0)}=\rp{2\pi}{n}=2\pi\sqrt{\rp{a^3}{GM}}\times
\sqrt{\rp{m_{\rm i}}{m_{\rm g}}}=2\pi\sqrt{\rp{a^3}{GM}}\times
\psi,\lb{due}\eqf where $G$, $M$ and $a$ are the Newtonian
gravitational constant, the mass of Earth and the satellite
semimajor axis, respectively. We have explicitly written the
square root of the ratio of the inertial to the (passive)
gravitational mass of satellite $\psi$. The quantity $\D n$
represents the various kind of perturbations, of gravitational and
non--gravitational origin, which affect $n$. For example, the even
zonal harmonic coefficients of the multipolar expansion of Earth
gravitational potential, called geopotential, induce secular
perturbations on $n$. The most important one is that due to the
first even zonal harmonic $J_2$ and, for a circular orbit with
eccentricity $e=0$, it is given by \eqi\D n^{(\ell=2)}_{\rm
obl}=-\rp{3}{4}J_2\left(\rp{R}{a}\right)^2(1-3\cos^2 i),\eqf where
$R$ is the Earth radius and $i$ is the inclination of the orbital
plane to the equator. Also the time--varying part of Earth
gravitational field should be considered, in principle, because
the Earth solid and ocean tides (Iorio 2001a) induce long--period
harmonic perturbations on $n$. For a given tidal line of frequency
$f$ the perturbations induced by the solid tides, which are the
most effective in affecting the orbits of a satellite, can be
written as \eqi \D n^{\ell}_{\rm tides
}=\sum_{m=0}^{\ell}\left(\rp{H_{\ell}^{m}}{R}\right)\left(\rp{R}{a}\right)^{\ell}k_{\ell
m}^{(0)}A_{\ell m}
\sum_{p=0}^{\ell}\sum_{q=-\infty}^{+\infty}F_{\ell mp}
\left[2(\ell +1)G_{\ell pq}-\rp{(1-e^2)}{e}\rp{dG_{\ell
pq}}{de}\right]\cos\gamma_{f\ell mpq},\eqf where $H_{\ell}^m$ are
the tidal heights, $k_{\ell m}^{(0)}$ are the Love numbers,
$A_{\ell m}=\sqrt{\rp{(2\ell+1)(\ell-m)!}{4\pi(\ell+m)!}}$,
$F_{\ell mp}(i)$ and $G_{\ell pq}(e)$ are the inclination and the
eccentricity functions (Kaula 1966), respectively, and
$\gamma_{f\ell mpq}$ is the frequency of the tidal perturbation
built up with the frequencies of the lunisolar variables and of
the satellite's orbital elements.

In order to test the WEP, we propose to measure, after many
revolutions, the difference of time spans which are multiple $N$
of the orbital periods $\D T_{N}\equiv T^{(2)}_N -T^{(1)}_N$ of a
couple of satellites of different composition orbiting the Earth
along circular orbits of almost same radius $a$ \eqi a^{(1)}\equiv
a,\ a^{(2)}=a+d.\lb{gluglu}\eqf The small difference $d$, which,
in principle, should be equal to zero, is due to the unavoidable
orbital injection errors. It can be made extremely small with a
rocket launcher of good quality: for example, at the beginning of
their mission the semimajor axes of the two GRACE spacecrafts were
different by an amount of just\footnote{See on the WEB
http://www.csr.utexas.edu/grace/newsletter/2002/august2002.html}
0.5 km.

The observable quantity in which we are interested is $\D \psi$ to
be measured after $N$ orbital revolutions, where
\eqi\psi^{(1)}\equiv \psi=1,\ \psi^{(2)}=\psi+\D\psi,\lb{vio}\eqf
with $\D\psi/\psi\ll 1$. The quantity $\D\psi$ can be expressed in
terms of the standard E${\rm \ddot o}$tv${\rm \ddot o}$s parameter
$\eta$. Indeed, the inertial mass of a body is composed by many
types of mass--energy: rest energy, electromagnetic energy,
weak--interaction energy, and so on. If one of these forms of
energy contributes to $m_{\rm g}$ differently than it does to
$m_{\rm i}$ we can put (Will 1993)\eqi m_{\rm g}=m_{\rm
i}+\sum_{A}\eta^A\rp{E^A}{c^2},\eqf where $E^A$ is the internal
energy of the body generated by the interaction $A$ and $\eta^A$
is a dimensionless parameter that measures the strength of the
violation of WEP induced by that interaction. Then \eqi
\rp{m^{(2)}_{\rm i}}{m^{(2)}_{\rm g}}-\rp{m^{(1)}_{\rm
i}}{m^{(1)}_{\rm
g}}\sim\sum_A\eta^A\left[\rp{E^A_{(1)}}{m^{(1)}_{\rm i
}c^2}-\rp{E^A_{(2)}}{m^{(2)}_{\rm i
}c^2}\right]\equiv\eta.\lb{eot}\eqf

From \rfr{eot} it can be obtained
\eqi\psi^{(2)}\equiv\sqrt{\rp{m^{(2)}_{\rm i}}{m^{(2)}_{\rm
g}}}=\sqrt{\rp{m^{(1)}_{\rm i}}{m^{(1)}_{\rm g
}}+\left[\rp{m^{(2)}_{\rm i}}{m^{(2)}_{\rm g }}-\rp{m^{(1)}_{\rm
i}}{m^{(1)}_{\rm g }}\right]}=\sqrt{\rp{m^{(1)}_{\rm
i}}{m^{(1)}_{\rm
g}}+\eta}\sim\psi^{(1)}+\rp{\eta}{2\psi^{(1)}},\eqf so that, from
\rfr{vio}, \eqi\Delta\psi=\rp{\eta}{2}.\eqf

From \rfrs{uno}{due} we can write \eqi \D T_N
=N2\pi\left[\rp{1}{n^{(2)}}-\rp{1}{n^{(1)}}+\rp{\D n^{(1)}_{\rm
obl} }{n^{(1)}}-\rp{\D n^{(2)}_{\rm obl}
}{n^{(2)}}\right].\lb{tre}\eqf In \rfr{tre} we consider  only the
gravitational even zonal perturbations due to Earth oblateness, as
usually done in orbital reduction programs in which a reference
orbit including the $J_2$ effects is adopted\footnote{This
approximation will be justified later.}. They can be summarized as
\eqi\D n_{\rm obl}=\sum
_{\ell=2}\left(\rp{R}{a}\right)^{\ell}\mathcal{G}_{\ell},\eqf
where the $\mathcal{G}_{\ell}=\mathcal{G}_{\ell}(i,\ e;\
J_{\ell})$ functions include the even zonal harmonics coefficients
$J_{\ell}$, the eccentricity $e$, the inclination angle $i$ and
some numerical constants. For example, for $\ell=2$ and $e=0$,
$\mathcal{G}_2=-\rp{3}{4}J_2(1-3\cos^2 i)$. By using the
expansions of \rfrs{gluglu}{vio} it is possible to solve \rfr{tre}
for\footnote{Here we have considered $i^{(1)}=i^{(2)}$. Moreover,
notice that if the two satellites would be inserted in
counter--rotating orbits, in $\D T_N$ it must be included also the
time shift $\D T^{\rm gm }_{N}\propto N\times 4\p\rp{J}{c^2M}$ due
to the general relativistic gravitomagnetic clock effect (Mashhoon
$et\ al$ 1999; 2001; Iorio $et\ al $ 2002) induced by the
off-diagonal components of the metric proportional to the proper
angular momentum $J$ of Earth. Indeed, it turns out (Iorio $et\
al$ 2002) that such time shift is independent of the ratio of the
inertial to the gravitational masses of the satellite. It would
act as a lower limit of the order of $N\times 10^{-7}$ s to $(\D
T_N)_{\rm exp}$ if it was detectable.} $\D\psi_N$
\eqi\D\psi_N=\rp{\rp{\sqrt{GM}}{N\pi}\D
T_N-(A+B)}{(A+C)+(B+D)},\lb{quattro}\eqf with \bea A&=& 3d\sqrt{a},\lb{cinquey}\\
B&=& \sum_{\ell=2}\mathcal{G}_{\ell} R^{\ell}\left[3\ell d^2
a^{-\left(\rp{1+2\ell}{2}\right)}+2\ell d
a^{\left(\rp{1-2\ell}{2}\right)} -3 d
a^{\left(\rp{1-2\ell}{2}\right)}\right],\\
C&=& 2\sqrt{a^3},\\
D&=& \sum_{\ell=2}\mathcal{G}_{\ell} R^{\ell}\left[-2
a^{\left(\rp{3-2\ell}{2}\right)} \right].\lb{sei}\eqfa
\Rfr{quattro}, together with \rfrs{cinquey}{sei}, allows to
evaluate the accuracy obtainable in measuring the quantity
$\D\psi_N$. The error in measuring the difference of multiples of
the orbital periods yields \eqi[\delta(\D \psi_N)]_{\D T_N
}=\rp{\sqrt{GM}}{N\pi}\rp{\delta(\D T_N)}{(A+C)+(B+D)},\eqf while
the error in Earth $GM$, which amounts to 8$\times 10^{11}$ cm$^3$
s$^{-2}$ (McCarthy 1996), yields \eqi[\delta(\D
\psi_N)]_{GM}=\rp{(\D T_N)_{\rm exp
}}{N2\pi\sqrt{GM}}\rp{\delta(GM)}{(A+C)+(B+D)}.\eqf The
uncertainty in Earth even zonal harmonics $J_{\ell}$, of which
$\delta J_2$ amounts to 7.9626$\times 10^{-11}$ (Lemoine $et\ al$
1998), has an impact given by \eqi[\delta(\D\psi_N)]_{J_{\ell}}
\leq\sum_{\ell=2}\left|\left\{\rp{-\rp{\partial B}{\partial
{J_{\ell}}}\times[(A+C)+(B+D)]+(A+B)\times\rp{\partial
(B+D)}{\partial
{J_{\ell}}}}{[(A+C)+(B+D)]^2}\right\}\right|\times\delta
J_{\ell}.\eqf The errors in $\D\psi_N$ due to the uncertainties in
$a$ and in $d$ are \bea [\delta(\D\psi_N)]_a
&=&\left|\left\{\rp{-\rp{\partial (A+B)}{\partial
a}\times[(A+C)+(B+D)]+(A+B)\times\rp{\partial
[(A+C)+(B+D)]}{\partial
a}}{[(A+C)+(B+D)]^2}\right\}\right|\times\delta
a,\\
{[\delta(\D\psi_N)]_d} &=&\left|\left\{\rp{-\rp{\partial
(A+B)}{\partial
d}\times(C+D)}{[(A+C)+(B+D)]^2}\right\}\right|\times\delta d.\eqfa
Notice that $[\delta(\D\psi_N)]_{J_{\ell}}$,
$[\delta(\D\psi_N)]_a$ and $[\delta(\D\psi_N)]_d$ do not depend
explicitly on the number of orbital revolutions $N$. In order to
calculate them we need the explicit expressions of the derivatives
with respect to $J_{\ell}$\bea \rp{\partial B}{\partial
{J_{\ell}}}&=& \left(\rp{\partial \mathcal{G}_{\ell}}{\partial
{J_{\ell}}}\right)R^{\ell}\left[3\ell d^2
a^{-\left(\rp{1+2\ell}{2}\right)}+2\ell d
a^{\left(\rp{1-2\ell}{2}\right)} -3 d
a^{\left(\rp{1-2\ell}{2}\right)}\right],\\
\rp{\partial D}{\partial {J_{\ell}}}&=& \left(\rp{\partial
\mathcal{G}_{\ell}}{\partial {J_{\ell}}} \right)R^{\ell}\left[-2
a^{\left(\rp{3-2\ell}{2}\right)} \right],\eea with respect to $a$
\bea
\rp{\partial A}{\partial a}&=& \rp{3d}{2\sqrt{a}},\lb{err1}\\
\rp{\partial B}{\partial a}&=& \sum_{\ell=2}\mathcal{G}_{\ell}
R^{\ell}\left[-3\rp{(1+2\ell)}{2}\ell d^2
a^{-\left(\rp{3+2\ell}{2}\right)}+(1-2\ell)\ell d
a^{-\left(\rp{1+2\ell}{2}\right)} -3\rp{(1-2\ell)}{2} d
a^{-\left(\rp{1+2\ell}{2}\right)}\right],\\
\rp{\partial C}{\partial a}&=& 3\sqrt{a},\\
\rp{\partial D}{\partial a}&=& \sum_{\ell=2}\mathcal{G}_{\ell}
R^{\ell}\left[-(3-2\ell)a^{\left(\rp{1-2\ell}{2}\right)}
\right],\lb{err2}\eqfa and those with respect to $d$
\bea \rp{\partial A}{\partial d}&=& 3\sqrt{a},\lb{cinque}\\
\rp{\partial B}{\partial d}&=& \sum_{\ell=2}\mathcal{G}_{\ell}
R^{\ell}\left[6\ell d a^{-\left(\rp{1+2\ell}{2}\right)}+2\ell
a^{\left(\rp{1-2\ell}{2}\right)} -3
a^{\left(\rp{1-2\ell}{2}\right)}\right].\eqfa

In order to fix the ideas, let us consider the orbit of the
proposed LARES laser--ranged satellite with $a= 12270$\ {\rm km},
$i$=70\ {\rm deg}. Let us assume (Peterson 1997) $d=5$ km.
%\delta a&=& 5\ {\rm mm},\\
%\delta d&=& 1\ {\rm cm}.
With these data we have, for $\ell=2$,
\begin{eqnarray}
[\delta(\Delta\psi_N)]_{\D T_N}&=& 7.3\times 10^{-5}\ {\rm s}^{-1}\times\rp{\delta(\D T_N)}{N},\lb{erro1}\\
{[\delta(\Delta\psi_N)]_{GM}} &=& 7.4\times 10^{-14}\ {\rm s}^{-1}\times\rp{(\D T_N)_{\rm exp}}{N},\\
{[\delta(\Delta\psi_N)]_{J_2}} &=& 8\times 10^{-15},\\
{[\delta(\Delta\psi_N)]_a} &=& 5\times 10^{-13}\ {\rm cm}^{-1}\times \delta a,\\
{[\delta(\Delta\psi_N)]_d} &=& 1\times 10^{-9}\ {\rm
cm}^{-1}\times \delta d.\lb{erro2}\end{eqnarray} The accuracy in
measuring the difference of the multiples of the orbital periods
of the satellites is a crucial factor in obtaining a high
precision in $\D\psi$. However, it could be possible, in
principle, to choose an observational time span covering a very
high number of orbital revolutions. It should be noted that $\d(\D
T_N)$ accounts for both the measurement errors and the
systematical errors induced by various gravitational and
non--gravitational aliasing phenomena. The latter ones play a very
important role in strongly limiting the possibility of measuring,
e.g., the gravitomagnetic clock effect: it amounts to 10$^{-7}$ s
after one orbital revolution for orbits with $e=i=0$ while in
(Iorio 2001b) it turned out that the systematic errors induced by
the present--day level of knowledge of the terrestrial
gravitational field are up to 2--3 order of magnitude larger. So,
it should not be unrealistic to consider $\d(\D T_N)_{\rm sys}\sim
10^{-4}-10^{-5}$ s. This would imply $[\d(\D\psi_N)]_{\D T_N}\sim
\rp{10^{-9}-10^{-10}}{N}$.

The error due to the difference $d$ in the semimajor axes of the
satellites, which turns out to be the major limiting factor,
cannot be reduced, in principle, by waiting for a sufficiently
high number of orbital revolutions because it is independent of
$N$. A small improvement could be obtained with the use of a
larger semimajor axis. A geostationary orbit with $a=42160$ km
would allow to get $[\delta(\Delta\psi_N)]_{\D T_N}= 1\times
10^{-5}\ {\rm s}^{-1}\times\rp{\delta(\D T_N)}{N}$ and
${[\delta(\Delta\psi_N)]_d} = 3\times 10^{-10}\ {\rm
cm}^{-1}\times \delta d.$ However, in this case, for a fixed time
span, we would have at our disposal a smaller $N$. In regard to
the systematic part of the error $\delta d$, it should be noted
that there are no secular or long--period perturbations of
gravitational origin on the semimajor axis of a satellite. The
non--gravitational perturbations could be reduced to a good level
by adopting the drag--free technology. In regard to $\delta d_{\rm
exp}$, it is important to note that in the present GRACE mission
(Davis $et\ al$ 1999), making use of a K/K$_{\rm a}$-band
 intersatellite link that provides dual one--way range
 measurements, changes in the distance of the two spacecrafts can
 be established with an accuracy of about 10$^{-2}$ cm or even better.

It is interesting to note that if we neglect in all calculations
$B$ and $D$ and the related derivatives, i.e., if we neglect the
effects of Earth oblateness, it turns out that the numerical
results of \rfrs{erro1}{erro2} do not change. It is very important
because it means that our choice of neglecting the contribution of
the non--gravitational perturbations in $\D n$ is $a\ posteriori$
correct. Indeed, the perturbing acceleration due to Earth $J_2$ on
LAGEOS is of the order of 10$^{-1}$ cm s$^{-2}$, while the impact
of the direct solar radiation pressure, which is the largest
non--gravitational perturbation on LAGEOS, amounts to 10$^{-7}$ cm
s$^{-2}$ (Milani $et\ al$ 1987). Moreover, these conclusions imply
that the errors in the inclination $i$, which enters the even
zonal harmonic perturbations due to the geopotential and the
non--gravitational perturbations, can be safely neglected, as done
here. The same considerations hold also for the tidal
perturbations: suffices it to say that the effects of the
18.6--year and the $K_1$ tides, which are the most powerful in
perturbing the satellite orbits, on a LAGEOS--type satellite are
six orders of magnitude smaller than those due to the static $J_2$
even zonal part of geopotential. Also the tiny general
relativistic gravitoelectric correction to the orbital period
induced by the Schwarzschild part of the metric\footnote{For a
circular orbit it is given by $T^{(0)}\times \Theta_{\rm
ge}=\psi\rp{3\pi\sqrt{GM a}}{c^2}$ (Mashhoon $et\ al$ 2001).},
which depends on $\sqrt{m_{\rm i}/m_{\rm g}}$, can be neglected
because for a LAGEOS--type satellite the disturbing acceleration
is of the order of 9$\times 10^{-8}$ cm s$^{-2}$ (Milani $et\ al$
1987).
%-----------------------------------------------------------------------------
\section{The longitude of the ascending node}
The longitude of the ascending node $\Omega$ is one of the best
accurately measured Keplerian orbital elements of Earth artificial
satellites. Then, we wish to examine if it would be possible to
use it in order to test the equivalence principle.

Let us recall that there are two kinds of long--period
perturbations on the node $\Omega$ of an Earth satellite. First,
the static oblateness of Earth induces a secular precession of
$\Omega$ through the even zonal harmonics of the geopotential.
Second, the time--varying part of Earth gravitational potential
induces tidal harmonic perturbations on $\Omega$ (Iorio 2001a).
Then, we can pose, by including the solid Earth tidal
perturbations
\eqi\dot\Omega=n\sum_{\ell=2}\left(\rp{R}{a}\right)^{\ell}\left[\mathcal{G}_{\ell}
+\sum_{m=0}^{\ell}\left(\rp{H_{\ell}^{m}}{R}\right)k_{\ell
m}^{(0)}A_{\ell m}
\sum_{p=0}^{\ell}\sum_{q=-\infty}^{+\infty}\rp{d F_{\ell mp}}{d i
}\rp{G_{\ell pq}}{\sin i\sqrt{1-e^2}}\cos\gamma_{f\ell
mpq}\right],\eqf where, for $\ell=2$ and $e=0$,
$\mathcal{G}_2=-\rp{3}{2}J_2\cos i$. By considering a couple of
satellites of different compositions freely orbiting along almost
identical orbits we could measure the difference of their secular
nodal rates $\Delta
\dot\Omega\equiv\dot\Omega^{(2)}-\dot\Omega^{(1)}$. In this case,
by posing $\xi\equiv\sqrt{\rp{m_{\rm g}}{m_{\rm i}}}$, the
violating parameter $\Delta\xi=-\rp{\eta}{2}$ can be expressed as
\eqi\Delta\xi=\rp{\rp{\Delta\dot\Omega}{\sqrt{GM}}+\mathcal{A}+\mathcal{B}}{\mathcal{C}-\mathcal{B}},\eqf
with\footnote{Here we neglect the non--gravitational perturbations
on the nodes. Contrary to, e.g., the perigees $\omega$, the nodes
are rather insensitive to such non--geodesic accelerations
(Lucchesi 2001; 2002). In regard to the Earth solid tides, they
have been neglected because their impact is several orders of
magnitude smaller. For example, the amplitude of the nodal rate
perturbation induced by the $K_1$ tide is five orders of magnitude
smaller than that due to the even zonal harmonic coefficient $J_2$
of the geopotential for a GPS orbit.} \bea
\mathcal{A}&=&\sum_{\ell=2}R^{\ell}\Delta\mathcal{G}_{\ell}a^{-\left(\rp{3+2\ell}{2}\right)},\\
\mathcal{B}&=&d\sum_{\ell=2}R^{\ell}\mathcal{G}_{\ell}\left(\rp{3+2\ell}{2}\right)a^{-\left(\rp{5+2\ell}{2}\right)},\\
\mathcal{C}&=&\sum_{\ell=2}R^{\ell}\mathcal{G}_{\ell}a^{-\left(\rp{3+2\ell}{2}\right)},\\\eqfa
where $\Delta
\mathcal{G}_{\ell}=\mathcal{G}^{(1)}_{\ell}-\mathcal{G}^{(2)}_{\ell}$
is the difference in the $\mathcal{G}_{\ell}$ functions of the two
satellites induced by the inclinations and the eccentricities. The
derivatives with respect to $J_{\ell}$, $a$ and $d$ are \bea
\rp{\partial\mathcal{A}}{\partial {J_{\ell}}}&=&\sum_{\ell=2}R^{\ell}\left(\rp{\partial\Delta\mathcal{G}_{\ell}}{\partial J_{\ell}}\right)a^{-\left(\rp{3+2\ell}{2}\right)},\\
\rp{\partial\mathcal{B}}{\partial {J_{\ell}}}&=&d\sum_{\ell=2}R^{\ell}\left(\rp{\partial\mathcal{G}_{\ell}}{\partial J_{\ell}}\right)\left(\rp{3+2\ell}{2}\right)a^{-\left(\rp{5+2\ell}{2}\right)},\\
\rp{\partial\mathcal{C}}{\partial {J_{\ell}}}&=&\sum_{\ell=2}R^{\ell}\left(\rp{\partial\mathcal{G}_{\ell}}{\partial J_{\ell}}\right)a^{-\left(\rp{3+2\ell}{2}\right)},\\
\rp{\partial\mathcal{A}}{\partial
a}&=&-\sum_{\ell=2}R^{\ell}\Delta\mathcal{G}_{\ell}
\left(\rp{3+2\ell}{2}\right)a^{-\left(\rp{5+2\ell}{2}\right)},\\
\rp{\partial\mathcal{B}}{\partial a}&=&
-d\sum_{\ell=2}R^{\ell}\mathcal{G}_{\ell}
\left(\rp{3+2\ell}{2}\right)\left(\rp{5+2\ell}{2}\right)a^{-\left(\rp{7+2\ell}{2}\right)},\\
\rp{\mathcal{\partial C}}{\partial a}&=&
-\sum_{\ell=2}R^{\ell}\mathcal{G}_{\ell}
\left(\rp{3+2\ell}{2}\right)a^{-\left(\rp{5+2\ell}{2}\right)},\\
\rp{\mathcal{\partial B}}{\partial
d}&=&\sum_{\ell=2}R^{\ell}\mathcal{G}_{\ell}
\left(\rp{3+2\ell}{2}\right)a^{-\left(\rp{5+2\ell}{2}\right)}.\eqfa
For $\ell=2$ and a GPS orbit, by assuming $a=26578$ km, $i=55$ deg
and $d=5$ km the errors in $\Delta\xi$ are\footnote{In the
calculations it turns out that the effect of $\D \mathcal{G}_{2}$
in $\mathcal{A}$ and $\rp{\partial \mathcal{A}}{\partial a}$, for
$\D i=1$ deg, can be neglected. On the other hand, with a good
quality rocket launcher it is possible to insert two spacecrafts
in the same orbital planes up to 10$^{-4}$ deg, as in the case of
the GRACE mission. See on the WEB
http://www.csr.utexas.edu/grace/newsletter/2002/august2002.html}
\begin{eqnarray}
[\delta(\Delta\xi)]_{\D\dot\Omega}&=& \rp{\delta(\D \dot\Omega)}{\sqrt{GM}(\mathcal{C}-\mathcal{B})}= 2\times 10^{-8}\ {(\rm mas/yr)}^{-1}\times\delta(\D \dot\Omega),\lb{errot1}\\
{[\delta(\Delta\xi)]_{GM}} &=& \rp{(\Delta\dot\Omega)_{\rm exp}}{2\sqrt{(GM)^3}(\mathcal{C}-\mathcal{B})}\times \delta(GM)=3\times 10^{-20}\ {(\rm deg/day)}^{-1}\times(\D\dot\Omega)_{\rm exp},\\
{[\delta(\Delta\xi)]_{J_2}} &=& \left|\left\{\rp{\rp{\partial
(\mathcal{A}+\mathcal{B})}{\partial {J_2}}
\times(\mathcal{C}-\mathcal{B})-\rp{\partial(\mathcal{C}-\mathcal{B})}{\partial
{J_2}}\times(\mathcal{A}+\mathcal{B})}
{(\mathcal{C}-\mathcal{B})^2}\right\}\right|\times \delta J_2=2\times 10^{-9},\\
{[\delta(\Delta\xi)]_a} &=& \left|\left\{\rp{\rp{\partial
(\mathcal{A}+\mathcal{B})}{\partial a}
\times(\mathcal{C}-\mathcal{B})-\rp{\partial(\mathcal{C}-\mathcal{B})}{\partial
a}\times(\mathcal{A}+\mathcal{B})}
{(\mathcal{C}-\mathcal{B})^2}\right\}\right|\times \delta a=2\times 10^{-13}\ {\rm cm}^{-1}\times \delta a,\\
{[\delta(\Delta\xi)]_d} &=& \left|\left\{\rp{\rp{\partial
\mathcal{B} }{\partial
d}\times(\mathcal{C}+\mathcal{A})}{(\mathcal{C}-\mathcal{B})^2}\right\}\right|\times
\delta d =1\times 10^{-9}\ {\rm cm}^{-1}\times \delta
d.\lb{errot2}\end{eqnarray} It can be noticed that the major
limiting factor is the term due to the error in the difference of
the nodal rates
$\delta(\D\dot\Omega)=\delta\dot\Omega^{(1)}+\delta\dot\Omega^{(2)}$.
Indeed, the experimental error in measuring the secular rate of
the node is of the order of 1 mas yr$^{-1}$. The systematic error
in $\dot\Omega$ due to the uncertainty on $J_2$ is, for a GPS
satellite, almost 3 mas yr$^{-1}$.  In the case of the error in
$d$, the same considerations as for the orbital periods hold.
%-----------------------------------------------------------------------------
\section{Conclusions}
In this paper we have shown that the comparison of the orbital
motions of a pair of artificial satellites of different
compositions moving along identical orbits in the gravitational
field of Earth in order to test the Weak Equivalence Principle is
not competitive with the already performed tests with torsion
balances on Earth and the Lunar Laser Ranging technique, and the
dedicated space--based missions STEP, GG, $\mu$SCOPE and SEE.

We have considered the orbital periods and the secular nodal
rates. The analysis of the orbital periods seems to yield more
precise measurements. The major limiting factor is represented by
the difference in the orbital radiuses induced by the unavoidable
orbital injection errors and the related uncertainty. By assuming
$\delta d\leq 1$ cm or less the achievable precision is of the
order of $10^{-10}$--$10^{-11}$.
%-----------------------------------------------------------------------------

\end{document}